\documentclass[12pt]{article}
\begin{document}
      \sloppy

\def\AFOUR{%
\setlength{\textheight}{9.0in}%
\setlength{\textwidth}{5.75in}%
\setlength{\topmargin}{-0.375in}%
\hoffset=-.5in%
\renewcommand{\baselinestretch}{1.17}%
\setlength{\parskip}{6pt plus 2pt}%
}
\AFOUR
\def\car{\mathop{\square}}
\def\carre#1#2{\raise 2pt\hbox{$\scriptstyle #1$}\car_{#2}}

\parindent=0pt
\makeatletter
\def\section{\@startsection {section}{1}{\z@}{-3.5ex plus -1ex minus
   -.2ex}{2.3ex plus .2ex}{\large\bf}}
\def\subsection{\@startsection{subsection}{2}{\z@}{-3.25ex plus -1ex minus
   -.2ex}{1.5ex plus .2ex}{\normalsize\bf}}
\makeatother
\makeatletter
\@addtoreset{equation}{section}
\renewcommand{\theequation}{\thesection.\arabic{equation}}
\makeatother

\renewcommand{\a}{\alpha}
\renewcommand{\b}{\beta}
\newcommand{\g}{\gamma}           \newcommand{\G}{\Gamma}
\renewcommand{\d}{\delta}         \newcommand{\D}{\Delta}
\newcommand{\e}{\varepsilon}
\newcommand{\la}{\lambda}        \newcommand{\LA}{\Lambda}
\newcommand{\m}{\mu}
\newcommand{\A}{\widehat{A}^{\star a}_{\mu}}
\newcommand{\Ar}{\widehat{A}^{\star a}_{\rho}}
\newcommand{\n}{\nu}
\newcommand{\om}{\omega}         \newcommand{\OM}{\Omega}
\newcommand{\p}{\psi}             \newcommand{\PS}{\Psi}
\renewcommand{\r}{\rho}
\newcommand{\s}{\sigma}           \renewcommand{\S}{\Sigma}
\newcommand{\f}{{\phi}}           \newcommand{\F}{{\Phi}}
\newcommand{\vf}{{\varphi}}
\newcommand{\y}{{\upsilon}}       \newcommand{\Y}{{\Upsilon}}
\newcommand{\z}{\zeta}

\renewcommand{\AA}{{\cal A}}
\newcommand{\BB}{{\cal B}}
\newcommand{\CC}{{\cal C}}
\newcommand{\DD}{{\cal D}}
\newcommand{\EE}{{\cal E}}
\newcommand{\FF}{{\cal F}}
\newcommand{\GG}{{\cal G}}
\newcommand{\HH}{{\cal H}}
\newcommand{\II}{{\cal I}}
\newcommand{\JJ}{{\cal J}}
\newcommand{\KK}{{\cal K}}
\newcommand{\LL}{{\cal L}}
\newcommand{\MM}{{\cal M}}
\newcommand{\NN}{{\cal N}}
\newcommand{\OO}{{\cal O}}
\newcommand{\PP}{{\cal P}}
\newcommand{\QQ}{{\cal Q}}
\renewcommand{\SS}{{\cal S}}
\newcommand{\RR}{{\cal R}}
\newcommand{\TT}{{\cal T}}
\newcommand{\UU}{{\cal U}}
\newcommand{\VV}{{\cal V}}
\newcommand{\WW}{{\cal W}}
\newcommand{\XX}{{\cal X}}
\newcommand{\YY}{{\cal Y}}
\newcommand{\ZZ}{{\cal Z}}

\newcommand{\ch}{\widehat{C}}
\newcommand{\gh}{\widehat{\gamma}}
\newcommand{\W}{W_{i}}
\newcommand{\na}{\nabla}
\newcommand{\xint}{\dint d^4x\;}
\newcommand{\sla}{\raise.15ex\hbox{$/$}\kern -.57em}
\newcommand{\Sla}{\raise.15ex\hbox{$/$}\kern -.70em}
\def\h{\hbar}
\def\Lp{\displaystyle{\biggl(}}
\def\Rp{\displaystyle{\biggr)}}
\def\LP{\displaystyle{\Biggl(}}
\def\RP{\displaystyle{\Biggr)}}
\newcommand{\lp}{\left(}\newcommand{\rp}{\right)}
\newcommand{\lc}{\left[}\newcommand{\rc}{\right]}
\newcommand{\lac}{\left\{}\newcommand{\rac}{\right\}}
\newcommand{\identity}{\bf 1\hspace{-0.4em}1}
\newcommand{\complex}{{\kern .1em {\raise .47ex
\hbox {$\scriptscriptstyle |$}}
      \kern -.4em {\rm C}}}
\newcommand{\real}{{{\rm I} \kern -.19em {\rm R}}}
\newcommand{\rational}{{\kern .1em {\raise .47ex
\hbox{$\scripscriptstyle |$}}
      \kern -.35em {\rm Q}}}
\renewcommand{\natural}{{\vrule height 1.6ex width
.05em depth 0ex \kern -.35em {\rm N}}}
\newcommand{\tint}{\int d^4 \! x \, }
\newcommand{\intg}{\int d^D \! x \, }
\newcommand{\intm}{\int_\MM}
\newcommand{\tr}{{\rm {Tr} \,}}
\newcommand{\half}{\dfrac{1}{2}}
\newcommand{\pa}{\partial}
\newcommand{\pad}[2]{{\frac{\partial #1}{\partial #2}}}
\newcommand{\fud}[2]{{\frac{\delta #1}{\delta #2}}}
\newcommand{\dpad}[2]{{\displaystyle{\frac{\partial #1}{\partial
#2}}}}
\newcommand{\dfud}[2]{{\displaystyle{\frac{\delta #1}{\delta #2}}}}
\newcommand{\dfrac}[2]{{\displaystyle{\frac{#1}{#2}}}}
\newcommand{\dsum}[2]{\displaystyle{\sum_{#1}^{#2}}}
\newcommand{\dint}{\displaystyle{\int}}
\newcommand{\eg}{{\em e.g.,\ }}
\newcommand{\Eg}{{\em E.g.,\ }}
\newcommand{\ie}{{{\em i.e.},\ }}
\newcommand{\Ie}{{\em I.e.,\ }}
\newcommand{\nb}{\noindent{\bf N.B.}\ }
\newcommand{\etal}{{\em et al.}}
\newcommand{\etc}{{\em etc.\ }}
\newcommand{\via}{{\em via\ }}
\newcommand{\cf}{{\em cf.\ }}
\newcommand{\twiddle}{\lower.9ex\rlap{$\kern -.1em\scriptstyle\sim$}}
\newcommand{\qed}{\vrule height 1.2ex width 0.5em}
\newcommand{\grad}{\nabla}
\newcommand{\bra}[1]{\left\langle {#1}\right|}
\newcommand{\ket}[1]{\left| {#1}\right\rangle}
\newcommand{\vev}[1]{\left\langle {#1}\right\rangle}

\newcommand{\equ}[1]{(\ref{#1})}
\newcommand{\eq}{\begin{equation}}
\newcommand{\eqn}[1]{\label{#1}\end{equation}}
\newcommand{\eea}{\end{eqnarray}}
\newcommand{\eqa}{\begin{eqnarray}}
\newcommand{\eqan}[1]{\label{#1}\end{eqnarray}}
\newcommand{\ba}{\begin{array}}
\newcommand{\ea}{\end{array}}
\newcommand{\eqac}{\begin{equation}\begin{array}{rcl}}
\newcommand{\eqacn}[1]{\end{array}\label{#1}\end{equation}}
\newcommand{\qq}{&\qquad &}
\renewcommand{\=}{&=&} 
\newcommand{\cb}{{\bar c}}
\newcommand{\mn}{{\m\n}}
\newcommand{\pic}{$\spadesuit\spadesuit$}
\newcommand{\?}{{\bf ???}}
\newcommand{\Tr }{\mbox{Tr}\ }
\newcommand{\adot}{{\dot\alpha}}
\newcommand{\bdot}{{\dot\beta}}
\newcommand{\gdot}{{\dot\gamma}}

\global\parskip=4pt
\titlepage  \noindent
{
   \noindent

\hfill GEF-TH-09/2006 


\vspace{2cm}

\noindent
{\bf
{\large General Solution Of Linear Vector Supersymmetry
}}

\vspace{.5cm}
\hrule

\vspace{2cm}

\noindent
{\bf 
Alberto Blasi and Nicola Maggiore}
\footnote{alberto.blasi@ge.infn.ge, nicola.maggiore@ge.infn.it,}

\noindent
{\footnotesize {\it
 Dipartimento di Fisica -- Universit\`a di Genova --
via Dodecaneso 33 -- I-16146 Genova -- Italy and INFN, Sezione di
Genova 
} }

\vspace{2cm}
\noindent
{\tt Abstract~:}
We give the general solution of the Ward identity for the linear vector 
supersymmetry which characterizes all topological models. Such solution, whose 
expression is quite compact and simple, greatly simplifies the study of 
theories displaying a supersymmetric algebraic structure, reducing to 
a few lines the proof of their possible finiteness. In particular, 
the cohomology technology usually involved for the quantum extension 
of these theories, is completely bypassed. The case of Chern-Simons 
theory is taken as an example.
\vfill\noindent
{\footnotesize {\tt Keywords:}
BRST Quantization,
Chern-Simons Theories,
Topological Field Theories,
Supersymmetric Gauge Theories.
\\
{\tt PACS Nos:} \\
03.70.+k Theory of Quantized Fields;\\
11.15.-q Gauge Field Theories;\\
11.10.Gh Renormalization;\\
11.10.Nx Noncommutative Field Theory.
}
\newpage
\begin{small}
\end{small}

\setcounter{footnote}{0}


\section{Introduction}

All topological field theories share that the energy-momentum tensor 
is not observable; for gauge field theories this feature is insured 
by a symmetry first discovered in~\cite{Delduc:1989ft}. We shall refer to it as 
the linear vector supersymmetry, which is encoded in a set $W_{\m}$ of 
anticommuting Ward operators, carrying a spacetime 
index $\m$. 

Like the BRS symmetry, the form of the vector supersymmetry is the 
same for any field theory displaying it. The reason for this is its 
relation with the gauge fixing term, as discussed in 
\cite{Maggiore:1991aa}, which, for any gauge field theory, is the 
same BRS cocycle. These two features of the vector supersymmetry, the 
fact of 
being shared by all topological field theories, and that of having the same 
functional form, render meaningful and interesting the task of 
finding its general solution. 

In this paper we will show that, first, a general solution for the 
vector supersymmetry does exist, and it can be cast in a very compact 
form, and, second, that such solution is so constraining to be 
sufficient, alone, to completely determine the quantum extension of 
any topological field theory.

Usually, in the BRS approach to renormalization of gauge field theories, 
the attention is focused on the BRS identity, and the 
ensuing algebraic discussion of the cohomology spaces; the vector 
supersymmetry is then imposed on the solutions of the BRS 
cohomological problem. This constrains some otherwise free parameters 
and helps in proving the finiteness of the theory. 

Here we would like to take the opposite attitude, and show that first 
discussing the general solution of the vector supersymmetry provides 
a more economical way towards the renormalization of topological 
gauge field theories and the discussion of observables in 
supersymmetric models.

Indeed, the general solution of the vector supersymmetry can be 
written in terms of 
${\cal W}\equiv\e^{\m_{1}\ldots\m_{d}}W_{\m_{1}}\ldots W_{\m_{d}}$, so that 
in $d$ spacetime dimensions, the cohomological analysis of the 
linearized BRS operator is shifted from the sector with power counting 
dimension $p$ and ghost charge $q$, to the sector with power counting 
dimension $p-d$ and ghost charge $q+d$. 

In all cases this results in a substantial simplification of the 
treatment. 

For twisted models, where the vector supersymmetry does exist but it 
is not linearly realized, the question remains open; for instance, it 
is known that the Lagrangian of twisted $N=2$ Super Yang-Mills can be 
cast in this form, but this is not a general result.

In order not to overload the notation, we shall illustrate the general 
procedure of solving the vector supersymmetry Ward identities in the case 
of the Chern-Simons theory with $d=3$, although the method can be 
straightforwardly extended to all other linear cases.

In section 2 we recall the defining functional equations for the 
Chern-Simons model, while the stability and anomaly problems are 
outlined in section 3. In Section 4 we give the general solution for 
the vector supersymmetry Ward identity, which we use in Section 5 to 
show in a few lines the finiteness of the theory. Section 6 is devoted 
to our concluding remarks.

\section{The Classical Model}

Let us consider, as an example, the topological three dimensional 
Chern-Simons (CS) theory, whose classical action is
\eq
S_{CS}=\frac{k}{2} \int d^{3}x\ \epsilon^{\mu\nu\rho} \left (
A^{a}_{\mu}\partial_{\nu}A^{a}_{\rho}
+ \frac{1}{3}f^{abc}
A^{a}_{\mu} A^{b}_{\nu} A^{c}_{\rho}
\right )\ ,
\eqn{2.1}
with the Landau gauge fixing
\eq
S_{gf} = \int d^{3}x\ \left (
b^{a}\partial A^{a} +\bar{c}^{a}\partial^{\mu}(D_{\mu}c)^{a}
\right)\ .
\eqn{2.2}
The fields $A^{a}_{\mu}(x)$, $c^{a}(x)$, $\bar{c}^{a}(x)$ and $b^{a}(x)$
represent gauge field, ghost, antighost and Lagrange multiplier 
(Nakanishi Lautrup field) respectively. They belong to the adjoint 
representation of a gauge group $G$, which we assume 
simple and 
compact. 
The $f^{abc}$ are the structure constants of the Lie algebra, $k$ is 
the coupling constant and finally the 
adjoint covariant derivative is defined as 
\eq
(D_{\mu}c)^{a} = \partial_{\mu}c^{a} + f^{abc}A^{b}_{\mu}c^{c}\ .
\eqn{2.3}

Besides the usual, nilpotent, BRS symmetry
\begin{eqnarray}
    s A^{a}_{\mu} &=& -(D_{\mu}c)^{a} \nonumber \\
    s c^{a} &=& +\frac{1}{2}f^{abc}c^{b}c^{c} \label{2.4}\\
    s \bar{c}^{a} &=& b^{a} \nonumber\\
    s b^{a} &=&  0\ ,\nonumber
\end{eqnarray}
under which $S_{CS}(A)$ and $S_{gf}(A,c,\bar{c},b)$ are separately 
invariant
\eq
s\ S_{CS}(A) = s\ S_{gf}(A,c,\bar{c},b) = 0\ ,
\eqn{2.5}
the gauge fixed action 
\eq
S= S_{CS} + S_{gf}
\eqn{2.6}
shows an additional, vector supersymmetry~\cite{Delduc:1989ft}
\eq
\delta_{\mu} S = 0\ ,
\eqn{2.7}
where 
\begin{eqnarray}
    \delta_{\mu} A^{a}_{\nu} &=& 
    \frac{1}{k}\epsilon_{\mu\nu\rho}\partial^{\rho}\bar{c}^{a}  \nonumber \\
    \delta_{\mu} c^{a} &=& -A^{a}_{\mu} \label{2.8}\\
    \delta_{\mu} \bar{c}^{a} &=& 0 \nonumber\\
    \delta_{\mu} b^{a} &=& \partial_{\mu}\bar{c}^{a}  \ ,\nonumber
\end{eqnarray}
whose origin is tightly related to the conservation of the 
    energy-momentum tensor~\cite{Maggiore:1991aa}, 
    which, for topological models, acquires 
    contribution from the gauge fixing term only, since
    \eq
    T_{\mu\nu} = \fud{S}{g^{\mu\nu}} = \fud{S_{gf}}{g^{\mu\nu}}\ ,
    \eqn{2.10}
    where $g^{\mu\nu}$ is the metric describing the manifold on which 
    the theory is built. Hence, the energy-momentum tensor is a BRS 
    cocycle
    \eq
    T_{\m\n}= s\Lambda_{\m\n}\ .
    \eqn{energymomentum}
 The vector supersymmetry is not a feature of the particular 
    model considered here, but its existence is a general property of 
    all topological field theories, of both Witten and 
    Schwartz type~\cite{Birmingham:1991ty}, built in the Landau gauge.

The set of constraints on the classical action $S$ is completed by
the (Landau) gauge condition
    \eq
    \fud{S}{b^{a}}=\partial A^{a}
    \eqn{2.13}
    and 
    the ghost condition
    \eq
    {\cal G}^{a}S=\int d^{3}x\ \left(
    \fud{}{c^{a}} + f^{abc}\bar{c}^{b}\fud{}{b^{c}} \right ) S=0\ ,
    \eqn{2.14}
    which is peculiar to any gauge field theory in the Landau 
    gauge~\cite{Blasi:1990xz}.

The nonlinearity of the BRS transformations \equ{2.4} renders necessary, 
in view of the quantum extension of the theory, the definition of the 
composite operators $sA^{a}_{\mu}$ and $sc^{a}$. This is done in 
a standard way in field theory, by coupling them to external sources  
(or antifields) 
$A^{\star a\mu}$ and 
$c^{\star a}$
by means of an additional term in the classical action :
\eq
S_{ext} = \int d^{3}x\ \left (
A^{\star a\mu} s A^{a}_{\mu} + c^{\star a}sc^{a}\right )\ .
\eqn{2.15}
The complete classical action 
\eq
\Sigma = S_{CS} + S_{gf} + S_{ext}
\eqn{2.16}
satisfies
\begin{enumerate}
    \item the Slavnov-Taylor (ST) identity
    \eq
{\cal S}(\Sigma) = \int d^{3}x \left (
\fud{\Sigma}{A^{*a\mu}} \fud{\S}{A^{a}_{\mu}} +
\fud{\Sigma}{c^{a*}} \fud{\S}{c^{a}}
+ b\fud{\S}{\bar{c}}
\right ) =0\ ,
\eqn{2.17}
which is the functional transcription of the BRS symmetry \equ{2.4};
\item the supersymmetric Ward identity assumes 
the same functional form in $any$ topological field theory 
\cite{Birmingham:1991ty,Birmingham:1991rh}.
\eq
W_{\mu}\Sigma=\Delta_{\mu}\ ,
\eqn{2.18}
where 
\eq
W_{\mu}= \int d^{3}x\
\left(
\frac{1}{k}\epsilon_{\mu\nu\rho}(\partial^{\rho}\bar{c}^{a} 
                              + A^{a*\rho})\fud{}{A^{a}_{\nu}} 
- A^{a}_{\mu}\fud{}{c^{a}} 
+\partial_{\mu}\bar{c}^{a}\fud{}{b^{a}} 
- c^{a*}\fud{}{A^{a*\mu}}
\right ) 
\eqn{2.19}
and $\Delta^{cl}_{\mu}$ is a breaking
\eq
\Delta^{cl}_{\mu} = \int d^{3}x\
\left ( 
\frac{1}{k}\epsilon_{\mu\nu\rho}A^{a*\nu}\partial^{\rho}b^{a}
-A^{a*\nu}\partial_{\mu}A^{a}_{\nu}
+c^{a*}\partial_{\mu}c^{a}
\right)\ .
\eqn{2.20}
As compared to \equ{2.7}, the vector supersymmetry on the 
complete classical action $\Sigma$, Eq.~\equ{2.18}, 
appears to be a broken symmetry. 
But, since $\Delta^{cl}_{\mu}$ is only linear in the 
quantum fields, such a breaking concerns only the classical 
theory~\cite{Piguet:1995er}. 
As we shall see, the Ward operator $W_{\mu}$ behaves as an exact 
symmetry for quantum objects such as counterterms and anomalies. This 
remark is crucial for what will follow;
\item the ghost equation, which, similarly to \equ{2.18}, on $\Sigma$ 
writes 
\eq
{\cal G}^{a}\Sigma = \Delta^{a}_{cl}\ ,
\eqn{2.21}
where again $\Delta_{cl}^{a}$ is a classical breaking
\eq
\Delta^{a}_{cl}= \int d^{3}x\ f^{abc}\left(
A^{b*\mu}A^{c}_{\mu}
- c^{b*}c^{c}
\right)\ ;
\eqn{2.22}
\item the antighost condition
\eq
\bar{\cal G}^{a}\Sigma=\left(
\partial^{\m}\fud{}{A^{a*\mu}}-\fud{}{\bar{c}^{a}}\right)\Sigma=0\ ,
\eqn{2.23}
in virtue of which $A^{a*\mu}$ and $\bar{c}^{a}$ appear in the 
combination
\eq
\A = A^{a*}_{\m} - \partial_{\mu}\bar{c}^{a}\ ;
\eqn{2.24}
\item
the Landau gauge condition \equ{2.13}, which formally remains the same
\eq
\fud{\S}{b^{a}} = \partial A^{a}\ .
\eqn{2.25}
\end{enumerate}

The following nonlinear algebra, good for a generic functional 
$\gamma$, with even Faddeev-Popov charge, holds:
\begin{eqnarray}
    B_{\g}{\cal S}(\g) &=& 0  \label{2.26} \\
    \{W_{\m},W_{\n}\} &=& 0 \label{2.27} \\
    W_{\m}{\cal S}(\g) + B_{\g}(W_{\m}\g-\D^{cl}_{\m}) &=& 
            {\cal P}_{\m}\g\ , \label{2.28} 
\end{eqnarray}
where $ B_{\S}$ is the linearized ST operator
\eq
B_{\S} = \int d^{3}x \left (
\fud{\Sigma}{A^{*a\mu}}\fud{}{A^{a}_{\m}} +
\fud{\S}{A^{a}_{\mu}} \fud{}{A^{a*\m}}        +
\fud{\Sigma}{c^{a*}} \fud{}{c^{a}} +
\fud{\S}{c^{a}} \fud{}{c^{a*}}
+ b\fud{\S}{\bar{c}}
\right ) =0
\eqn{2.29}
and ${\cal P}_{\m}$ is the translation Ward operator
\eq
{\cal P}_{\m}=\sum_{\mbox{all fields}\ \Phi}\int d^{3}x\ 
(\partial_{\m}\Phi)\fud{}{\Phi}\ .
\eqn{2.30}
If the functional $\g$ satisfies the ST identity \equ{2.17}
and the Ward identity \equ{2.18}, the algebra for the linear operator 
$B_{\S}$ and $W_{\m}$ is
\begin{eqnarray}
    (B_{\S})^{2} &=& 0 \nonumber \\
    \{W_{\m},W_{\n}\} &=& 0 \label{2.31} \\
    \{B_{\S},W_{\m}\} &=& {\cal P}_{\m}\ . \nonumber 
\end{eqnarray}
Notice that the introduction of the external sources $A^{\star a\mu}$ and 
    $c^{\star a}$ allows to put off shell the algebra \cite{Delduc:1989ft}. 
    Under this 
respect, the external sources, which naturally enter the game in the 
BRS formalism, play exactly the role of the auxiliary fields of the 
ordinary 
supersymmetry~\cite{West:1990tg}.

\section{The Quantum Theory}

The quantum theory is well defined once we have proved that the 
divergences can be reabsorbed through a redefinition of fields and 
parameters of the theory, order by order in perturbation theory, and 
once we have been able to show the absence of quantum obstructions 
for the symmetries defining the theory. 

In other words, the quantum 
extension of a generic field theory goes through the determination of 
the counterterm and of the proof of the absence of anomalies.

Topological fields theories, like the one we are considering in this 
paper, are an example of $finite$ field theories~\cite{Maggiore:1991aa}. 
This means that no
counterterm should be allowed, {\it i.e.} a finite model does not display any 
divergence which needs to be reabsorbed. 

As far as we know, the only 
field theories showing this remarkable property are topological 
    models and supersymmetric gauge field theories~\cite{West:1990tg}. 
    It is not by chance, 
indeed, that topological fields theories are characterized, through 
the presence of the linear vector supersymmetry $W_{\m}$, by a typically 
supersymmetric algebra 
like~\equ{2.31}~\cite{Maggiore:1991aa,Guadagnini:1990aw,Guadagnini:1990br,
Maggiore:1992ug,Blasi:1992hq,Birmingham:1991rh}. 
In some cases, ordinary gauge 
field theories and topological quantum field theories even coincide, being 
possible to pass from one to the other by means of a twisting 
procedure~\cite{Witten:1988ze}. This is the case, for instance, of 
N=2 Super Yang Mills 
(SYM) theory and topological Yang-Mills (TYM) 
theory~\cite{Fucito:1997xm,Blasi:2000qw}, where the vector 
supersymmetry is not linearly realized, and our method cannot be 
applied straightforwardly.

For CS theory, finiteness has been proved in~\cite{Blasi:1989mw} and 
in~\cite{Delduc:1990je}, 
independently, and following two different approaches. In this paper we give 
an alternative proof, and, even if we believe that the method 
presented here gives a new, particularly quick and economic 
demonstration of 
the finiteness of the theory, we stress again that our main aim is to 
point out the power of supersymmetry, and of its general solution 
which will be given in the next section. Supersymmetry alone turns out 
to be sufficient to completely determine the quantum extension of 
these theories.

A generic local integrated functional $X^{p}_{q}$ of power counting 
dimension $p$ and ghost number $q$ which satisfies the algebraic 
constraints beyond the classical level obeys 
\begin{eqnarray}
\fud{X^{p}_{q}}{b^{a}(x)} &=& 0 \label{3.2} \\
\bar{\cal G}^{a}(x) X^{p}_{q} &=& 0 \label{3.3} \\
{\cal G}^{a}X^{p}_{q} &=& 0 \label{3.4} \\
W_{\m}X^{p}_{q} &=& 0 \label{3.5} \\
B_{\S}X^{p}_{q} &=& 0\ . \label{3.6} 
\end{eqnarray}
It is well known \cite{Piguet:1995er} that the linear symmetries \equ{3.2}, 
\equ{3.3} and \equ{3.4} can be extended to the quantum level, and 
therefore we have to analyze \equ{3.5} and \equ{3.6}, the relevant 
cases for the quantum extension of the theory being $X^{3}_{0}$ 
for the counterterm and $X^{3}_{1}$ for the anomaly.

\section{General Solution Of Supersymmetry Constraint}

The aim of this section, is to study the most general solution of the 
supersymmetry constraint \equ{3.5}
\eq
W_{\m}X^{p}_{q}=0\ .
\eqn{4.1}
The interest for solving this equation is evident, since
it must be satisfied by both counterterm and gauge anomaly, in 
topological and (twisted) supersymmetric field theories as well. 
This, together with the fact that the form of $W_{\m}$ is independent 
from the particular model considered, being 
related only to the gauge fixing term, according to \equ{2.10}, 
justifies our effort for finding its general solution.

In the functional space defined by  the gauge condition \equ{3.2}, 
the antighost equation 
\equ{3.3} 
and the ghost equation \equ{3.4}, the Ward operator $W_{\m}$ reads
\eq
W_{\m}=\int d^{3}x\ \left(
\frac{1}{k}\e_{\m\n\r}\hat{A}^{a*\r}\fud{}{A^{a}_{\n}}
- A^{a}_{\m}\fud{}{c^{a}}
+ c^{a*}\fud{}{\hat{A}^{a*\m}}
\right)\ ,
\eqn{4.2}
and it satisfies the algebraic relation \equ{2.27}.

It is convenient to make the Lorentz index explicit:
\begin{eqnarray}
W_{1} &=& \int d^{3}x\ \left(
\frac{1}{k}\hat{A}^{a*}_{3}\fud{}{A^{a}_{2}}
- \frac{1}{k}\hat{A}^{a*}_{2}\fud{}{A^{a}_{3}}
- A^{a}_{1}\fud{}{c^{a}}
+c^{a*}\fud{}{\hat{A}^{a*}_{1}}
\right ) \label{4.3} \\
W_{2} &=& \int d^{3}x\ \left(
-\frac{1}{k}\hat{A}^{a*}_{3}\fud{}{A^{a}_{1}}
+ \frac{1}{k}\hat{A}^{a*}_{1}\fud{}{A^{a}_{3}}
- A^{a}_{2}\fud{}{c^{a}}
+c^{a*}\fud{}{\hat{A}^{a*}_{2}}
\right ) \label{4.4} \\
W_{3} &=& \int d^{3}x\ \left(
\frac{1}{k}\hat{A}^{a*}_{2}\fud{}{A^{a}_{1}}
- \frac{1}{k}\hat{A}^{a*}_{1}\fud{}{A^{a}_{2}}
- A^{a}_{3}\fud{}{c^{a}}
+c^{a*}\fud{}{\hat{A}^{a*}_{3}}
\right ) \label{4.5} 
\end{eqnarray}
The algebra between the $W$'s is
\eq
W_{1}^{2}=W_{2}^{2}=W_{3}^{2}=0
\eqn{4.6}
\eq
\{W_{1},W_{2}\} = \{W_{1},W_{3}\} = \{W_{2},W_{3}\} =0\ .
\eqn{4.7}
In particular, the three operators $W$ are nilpotent. Now, from their 
explicit expressions, we see that all fields forming the functional 
space on which the $W$'s act, appear as BRS doublets \cite{Piguet:1995er}, 
{\it i.e.} are of 
the type:
\eq
W_{i}\phi=\psi\ \ ,\ \ W_{i}\psi=0\ \ \ i=1,2,3
\eqn{4.8}
or, equivalently
\eq
\{W_{i},W^{\dagger}_{j}\} = \d_{ij}{\cal N}\ ,
\eqn{4.9}
where the $W^{\dagger}_{i}$ are the functional adjoints of $W_{i}$
\begin{eqnarray}
W^{\dagger}_{1} &=& \int d^{3}x\ \left(
kA^{a}_{2}\fud{}{\hat{A}^{a*}_{3}}
- kA^{a}_{3}\fud{}{\hat{A}^{a*}_{2}}
- c^{a}\fud{}{A^{a}_{1}}
+\hat{A}^{a*}_{1}\fud{}{c^{a*}}
\right ) \label{4.10} \\
W^{\dagger}_{2} &=& \int d^{3}x\ \left(
-kA^{a}_{1}\fud{}{\hat{A}^{a*}_{3}}
+ kA^{a}_{3}\fud{}{\hat{A}^{a*}_{1}}
- c^{a}\fud{}{A^{a}_{2}}
+\hat{A}^{a*}_{2}\fud{}{c^{a*}}
\right ) \label{4.11} \\
W^{\dagger}_{3} &=& \int d^{3}x\ \left(
kA^{a}_{1}\fud{}{\hat{A}^{a*}_{2}}
- kA^{a}_{2}\fud{}{\hat{A}^{a*}_{1}}
- c^{a}\fud{}{A^{a}_{3}}
+\hat{A}^{a*}_{3}\fud{}{c^{a*}}
\right ) \label{4.12} 
\end{eqnarray}
and the operator ${\cal N}$
\eq
{\cal N}=\int d^{3}x\
\sum_{\mbox{all fields $\Phi$}}
\Phi\fud{}{\Phi}
\eqn{4.13}
counts  the number $n_{\Phi}$ of fields $\Phi$
appearing in a generic functional ${\cal F}(\Phi)$
\eq
{\cal N}{\cal F}(\Phi) =
\sum_{\mbox{all fields $\Phi$}} n_{\Phi}\ {\cal F}(\Phi)
\equiv N\ {\cal F}(\Phi)\ .
\eqn{4.14}
According to a well known theorem concerning nilpotent operators 
\cite{Piguet:1995er}, fields transforming as BRS doublets, do not belong to the 
cohomology. This means that the cohomology of each $W_{i}$ is empty, 
since all fields are organized in BRS doublets, and that the 
solutions of the equations
\eq
W_{i}X^{p}_{q}=0
\eqn{4.15}
can be written as
\eq
X^{p}_{q}= W_{i}X^{p-1}_{q+1}\ ,
\eqn{4.16}
where we took into account the fact that the operators $W$ raise by one unit the mass 
dimensions and lower the ghost number by the same amount.

But we can go further. We are looking, indeed, for the most general 
$X^{p}_{q}$ which satisfies the set of equations
\eq
W_{1}X^{p}_{q}=W_{2}X^{p}_{q}=W_{3}X^{p}_{q}=0\ .
\eqn{4.17}
Since the cohomology of each $W_{i}$ is empty, we have
\eq
X^{p}_{q} = W_{1}\bar{X}^{p-1}_{q+1} 
= W_{2}\widehat{X}^{p-1}_{q+1} 
= W_{3}\widetilde{X}^{p-1}_{q+1}\ , 
\eqn{4.18}
where $\bar{X}^{p-1}_{q+1}$, $\widehat{X}^{p-1}_{q+1}$ and 
$\widetilde{X}^{p-1}_{q+1}$ 
are generic functionals.

Multiplying both sides of the second identity in \equ{4.18} by 
$W_{1}^{\dagger}$, we have
\eq
W_{1}^{\dagger} W_{1}\bar{X}^{p-1}_{q+1}  =
W_{1}^{\dagger} W_{2}\widehat{X}^{p-1}_{q+1}\ ,
\eqn{4.19}
and, using \equ{4.9}, we get
\eq
{\cal N}\bar{X}^{p-1}_{q+1}
- W_{1} W_{1}^{\dagger}\bar{X}^{p-1}_{q+1}
= - W_{2}W_{1}^{\dagger} \widehat{X}^{p-1}_{q+1}\ .
\eqn{4.20}
Thanks to \equ{4.14}, we can write
\eq
\bar{X}^{p-1}_{q+1}=\frac{1}{N_{1}}\left (
W_{1} W_{1}^{\dagger}\bar{X}^{p-1}_{q+1}
- W_{2}W_{1}^{\dagger} \widehat{X}^{p-1}_{q+1}
\right)\ ,
\eqn{4.21}
where $N_{1}$ is the number of fields appearing in $\bar{X}^{p-1}_{q+1}$.
Similarly, from the third identity in \equ{4.18}, we have
\eq
W_{2}^{\dagger} W_{2}\widehat{X}^{p-1}_{q+1}  =
W_{2}^{\dagger} W_{3}\widetilde{X}^{p-1}_{q+1}\ ,
\eqn{4.22}
hence
\eq
{\cal N}\widehat{X}^{p-1}_{q+1}
- W_{2} W_{2}^{\dagger}\widehat{X}^{p-1}_{q+1}
= - W_{3}W_{2}^{\dagger} \widetilde{X}^{p-1}_{q+1}\ ,
\eqn{4.23}
{\it i.e.}
\eq
\widehat{X}^{p-1}_{q+1}=\frac{1}{N_{2}}\left (
W_{2} W_{2}^{\dagger}\widehat{X}^{p-1}_{q+1}
- W_{3}W_{2}^{\dagger} \widetilde{X}^{p-1}_{q+1}
\right)\ ,
\eqn{4.24}
where $N_{2}$ is the number of fields appearing in $\widehat{X}^{p-1}_{q+1}$.

Substituting \equ{4.24} in \equ{4.21}, we get
\begin{eqnarray}
\bar{X}^{p-1}_{q+1} &=&
\frac{1}{N_{1}}W_{1}W_{1}^{\dagger}\bar{X}^{p-1}_{q+1}
-\frac{1}{N_{1}N_{2}}W_{2}W_{1}^{\dagger}
\left(W_{2}W_{2}^{\dagger}\widehat{X}^{p-1}_{q+1}
-W_{3}W_{2}^{\dagger}\widetilde{X}^{p-1}_{q+1}
\right) \nonumber \\
&=& 
\frac{1}{N_{1}}W_{1}W_{1}^{\dagger}\bar{X}^{p-1}_{q+1}
+
\frac{1}{N_{1}N_{2}}W_{2}W_{1}^{\dagger}
W_{3}W_{2}^{\dagger}\widetilde{X}^{p-1}_{q+1}\ , \label{4.25}
\end{eqnarray}
because of the nilpotency of $W_{2}$ and of the algebra \equ{4.9}.

We are now able to write, from \equ{4.18}
\eq
X^{p}_{q} =
\frac{1}{N_{1}N_{2}}
W_{1}W_{2}W_{1}^{\dagger}W_{3}W_{2}^{\dagger}
\widetilde{X}^{p-1}_{q+1}\ .
\eqn{4.26}

Therefore, the most general solution of the equations \equ{4.17} is, by 
construction
\eq
X^{p}_{q} =
W_{1}W_{2}W_{3}
Y^{p-3}_{q+3}\ .
\eqn{4.27}
In covariant notations our result writes
\eq
W_{\m}X^{p}_{q}=0
\rightleftharpoons
X^{p}_{q}=\e^{\a\b\g}W_{\a}W_{\b}W_{\g}
Y^{p-3}_{q+3}\ ,
\eqn{4.28}
where $Y^{p-3}_{q+3}$ is a generic functional with dimensions $p-3$ 
and ghost number $q+3$. 

Although our proof has been given in three dimensions, the procedure 
we employed holds for any topological field theory in $d$-spacetime, and hence we can write the most general 
solution of the set of $d$ equations
\eq
W_{\m}X^{p}_{q} = 0\ ,
\eqn{4.29}
as
\eq
X^{p}_{q} = {\cal W}X^{p-d}_{q+d}\ ,
\eqn{4.30}
where
\eq
{\cal W} \equiv
\e^{\m_{1}\ldots\m_{d}}
W_{\m_{1}}\ldots W_{\m_{d}}
\ .
\eqn{4.31}
The fact that the operators $W_{\m}$ carry both dimensions and 
Faddeev-Popov charge, is quite constraining, and the consequences on 
counterterm and anomalies are spectacular, as we shall see in the 
next section.

We finally stress that, due to the algebra \equ{2.27}, a functional 
written as \equ{4.30} trivially satisfies the supersymmetry equation 
\equ{4.29}. What we have shown here, is that no other solutions exist.

\section{Supersymmetry And Finiteness}

The cohomology equations \equ{3.6}, written for the counterterm and for 
the anomaly, are usually considered, and generally are, as the hearth 
of the quantum determination of the theory, and of the algebraic 
renormalization.

Smart and subtle techniques have been developed for solving them, 
like the method of spectral sequences~\cite{Dixon:1991wi} or the so called 
``russian formula''~\cite{Zumino:1983ew}, 
and useful theorems exist, which
give information on their solutions~\cite{Piguet:1995er}. Brute force, on the 
other hand, often required a lot of time and hard work to obtain 
important results concerning the renormalization (or non 
renormalization) of the theory. What we want to show in this paper, 
is that for the large number of theories which exhibit the vector 
supersymmetry, the way to gain quantum theory is much shorter 
and bypasses the cohomology technology.

Let us begin with the counterterm $X^{3}_{0}$, which must satisfy the 
supersymmetry condition \equ{3.5}. 

We have just shown that the most general solution is
\eq
X^{3}_{0} = {\cal W} X^{0}_{3}\ ,
\eqn{5.1}
where $X^{0}_{3}$ is an integrated local functional  with vanishing 
power counting dimension and ghost number three. The only 
possibility is
\eq
X^{0}_{3} = \a
\int d^{3}x\
f^{abc} c^{a}c^{b}c^{c}\ ,
\eqn{5.2}
where $\a$ is a coefficient. This only candidate for the counterterm 
is ruled out by the ghost condition \equ{3.4}, because of the 
algebraic condition
\eq
\{{\cal G}^{a},{\cal W}\} =0\ ,
\eqn{5.3}
which can be easily verified.

Hence, 
\eq
X^{3}_{0}=0\ ,
\eqn{5.4}
which proves the finiteness of the theory.

The proof of absence of anomalies is even shorter. Besides the 
Wess-Zumino consistency condition \equ{3.6}, the gauge anomaly 
$X^{3}_{1}$ 
must satisfy the supersymmetry equation \equ{3.5}, whose most general 
solution is
\eq
X^{3}_{1}={\cal W}X^{0}_{4}\ ,
\eqn{5.5}
where $X^{0}_{4}$ is a generic integrated local functional functional 
with mass dimension zero and Faddeev-Popov charge four.

The only possibility would be
\eq
X^{0}_{4} = \int d^{3}x\
T^{abcd}c^{a}c^{b}c^{c}c^{d}\ ,
\eqn{5.6}
but such candidate is not there, simply because an invariant tensor 
$T^{abcd}$, which must be completely antisymmetric in its four color 
indices, cannot even be written~\cite{MacFarlane:1968vc}.

That's the reason why
\eq
X^{3}_{1}=0\ ,
\eqn{5.7}
and therefore no anomalies exist for any of the symmetries defining 
the model.

Notice that, to show the results \equ{5.4} and \equ{5.7}, no use has been done of the 
left conditions on the counterterm and on the anomaly, which are by far the 
toughest ones to analyze, {\it i.e.} the ST conditions \equ{3.6}. As 
promised.

\section{An Example: Absence Of Anomalies In Noncommutative Chern-Simons Theory}

Besides finiteness of Chern-Simons theory, as another, original, 
application of our result \equ{4.30}, in this section we face 
the problem of anomalies for noncommutative Chern-Simons (NCCS) theory.

As it is well known, a noncommutative extension of a generic 
quantum field theory is obtained through the substitution of the 
ordinary product with the Groenewald-Moyal 
\cite{Douglas:2001ba,Szabo:2001kg}:
\eq
\phi(x)\psi(x) \longrightarrow \phi(x)*\psi(x) \equiv
\lim_{y\rightarrow x}\
\exp(\frac{i}{2}\theta^{\mu\nu}\partial^{x}_{\mu}\partial^{y}_{\nu})\
\phi(x)\psi(y)\ ,
\eqn{6.1}
where $\theta^{\mu\nu}$ is a rank-two antisymmetric matrix which 
controls the noncommutative nature of spacetime coordinates
\eq
[x^{\mu},x^{\nu}]=i\theta^{\mu\nu}\ .
\eqn{6.2}

The NCCS theory, at $O(\theta^{2})$,  reads
\begin{eqnarray}
S_{NCCS} &=& \frac{k}{2}\ \Tr \int d^{3}x\ \epsilon^{\mu\nu\rho} \left (
A_{\mu}*\partial_{\nu}A_{\rho}
- i\frac{2}{3}
A_{\mu} * A_{\nu} * A_{\rho}
\right ) \nonumber \\
&=&
S_{CS} + \theta^{\a\b}\int d^{3}x\
\left (
\frac{1}{12}\e^{\m\n\r}d^{abc}
(\partial_{\a}A^{a}_{\m}) (\partial_{\b}A^{b}_{\n}) A^{c}_{\r}
\right) + \label{6.3} \\
&&+
\theta^{\a\b}\theta^{\g\d} \int d^{3}x\
\left(
-\frac{1}{48} \e^{\m\n\r}f^{abc}
(\partial_{\a\g}A^{a}_{\m})
(\partial_{\b\d}A^{b}_{\n})
A^{c}_{\r}
\right)\ . \nonumber
\end{eqnarray}

The noncommutative gauge fixing term, , at $O(\theta^{2})$, is
\begin{eqnarray}
S_{NCgf} &=& s^{(\theta)}\ \Tr \int d^{3}x\ \bar{c}*\partial^{\m}A_{\m}
\nonumber \\
&=&
S_{gf} + \theta^{\a\b}\int d^{3}x\
\left (
-\frac{1}{2}d^{abc}
(\partial^{\m}\bar{c}^{a})
(\partial_{\a}A^{b}_{\m}) 
(\partial_{\b}c^{c})
\right )+ \label{6.4} \\
&&+
\theta^{\a\b}\theta^{\g\d} \int d^{3}x\
\left (
\frac{1}{8} f^{abc}
\partial^{\m}\bar{c}^{a}
(\partial_{\a\g}A^{b}_{\m})
(\partial_{\b\d}c^{c})
\right)\ , \nonumber
\end{eqnarray}
where $s^{(\theta)}$ is the noncommutative extension of the ordinary 
BRS operator \equ{2.4}, always at $O(\theta^{2})$:
\begin{eqnarray}
    s^{(\theta)} A^{a}_{\mu} &=& s A^{a}_{\mu}
    -\frac{1}{2}\theta^{\a\b}d^{abc}\partial_{\a}A^{b}_{\m}\partial_{\b}c^{c}
    +\frac{1}{8}\theta^{\a\b}\theta^{\g\d}f^{abc}\partial_{\a\g}A^{b}_{\m}\partial_{\b\d}c^{c}
    \nonumber \\
    s^{(\theta)} c^{a} &=& s c^{a}
    +\frac{1}{4}\theta^{\a\b}d^{abc}\partial_{\a}c^{b}\partial_{\b}c^{c}
    -\frac{1}{16}\theta^{\a\b}\theta^{\g\d}f^{abc}\partial_{\a\g}c^{b}
    \partial_{\b\d}c^{c}\label{6.5}\\
    s^{(\theta)} \bar{c}^{a} &=&  s \bar{c}^{a} \nonumber\\
    s^{(\theta)} b^{a} &=&   s b^{a}\ .\nonumber
\end{eqnarray}
The noncommutative action \equ{6.3}, although quite different from the 
ordinary one \equ{2.1}, is still invariant under 
supersymmetry, which, being a linear transformation, is not modified 
by $\theta$:
\eq
\d_{\m}\
\left (
S_{NCCS} + S_{NCgf}
\right)
=0\ .
\eqn{6.6}

The introduction of the noncommutative parameter deeply alters the 
nature of any theory, firstly because it breaks Lorentz invariance. 
In addition, it has been shown that, at least in the two-dimensional 
noncommutative BF model \cite{Blasi:2005bk,Blasi:2005vf}, the 
$\theta$-deformed theory is not stable under radiative corrections, 
since the presence of $\theta$ opens a sector for counterterms which 
are not present at the classical level, and which cannot be traced back to 
a Groenewald-Moyal product, which turns out to be unstable as well.

As an open question remains the issue of anomalies, since there are 
no results concerning the cohomology of the BRS operator in presence 
of a $\theta$-parameter \cite{Picariello:2001mu}. 
The fact that $\theta$ has negative 
mass dimensions, renders the algebraic analysis overwhelmingly heavy, 
and, in practice, results at higher orders in $\theta$ turn out to be 
out of reach.

Our result considerably simplifies the analysis, and the algebraic 
treatment, even at higher orders in $\theta$, is still 
feasible.

The 3D gauge anomaly is a local integrated functional 
${\cal A}^{(\theta)3}_{1}$ with mass dimensions $+3$ and ghost charge $+1$, which 
satisfies all the symmetries defining the classical action, namely 
the ghost equation
\eq
 {\cal G}^{a}{\cal A}^{(\theta)3}_{1} =0\ ,
\eqn{6.7}
the supersymmetry condition
\eq
W_{\m}{\cal A}^{(\theta)3}_{1}=0\ ,
\eqn{6.8}
and, finally, the Slavnov-Taylor condition
\eq
B_{\Sigma}^{(\theta)}{\cal A}^{(\theta)3}_{1}=0\ .
\eqn{6.9}
Concerning the last constraint, namely the Wess-Zumino condition, in 
order that the functional ${\cal A}^{(\theta)3}_{1}$ is a true 
anomaly, it must 
belong to the cohomology of $B_{\Sigma}^{(\theta)}$, namely:
\eq
{\cal A}^{(\theta)3}_{1} \neq B_{\Sigma}^{(\theta)}{\cal 
A}^{(\theta)3}_{0}.
\eqn{6.10}
In the noncommutative case, ${\cal A}^{(\theta)3}_{1}$ can be expressed as a 
power series in $\theta$\footnote{It can be easily proved that 
in 3D
no non-analytical sector exists in $\theta$.}:
\eq
{\cal A}^{(\theta)3}_{1}= 
{\cal A}^{3}_{1} +
\sum_{n=1}^{\infty}
\theta_{\a_{1}\b_{1}}..\theta_{\a_{n}\b_{n}}
({\cal A}^{3+2n}_{1})^{\a_{1}\b_{1}..\a_{n}\b_{n}}\ .
\eqn{6.11}

Since the linear operator $W_{\m}$ is $\theta$-independent, the 
equation \equ{6.8} must hold order by order in $\theta$:
\eq
W_{\m} ({\cal A}^{3+2n}_{1})^{\a_{1}\b_{1}..\a_{n}\b_{n}} =0\ ,
\eqn{6.12}
which is the equation whose general solution we give in this paper, 
according to which
\eq
({\cal A}^{3+2n}_{1})^{\a_{1}\b_{1}..\a_{n}\b_{n}}
=
{\cal W}\
({\cal A}^{2n}_{4})^{\a_{1}\b_{1}..\a_{n}\b_{n}}\ ,
\eqn{6.13}
{\it i.e.}, to find 
candidates for the anomaly  at the order  $\theta^{n}$, 
we must study the cohomology of the Slavnov-Taylor operator in the 
space of local integrated functionals with mass dimension $2n$, 
ghost charge $+4$, with $2n$ Lorentz indices. This greatly reduces the 
number of possibilities.

Indeed:

The (commutative) order $\theta^{0}$ has already been treated in the 
previous section, with the immediate outcome that no commutative 
anomaly exists.

At order $O(\theta)$: 
\eq
({\cal A}^{5}_{1})^{\a\b}
=
{\cal W}\
({\cal A}^{2}_{4})^{\a\b}\ ,
\eqn{6.14}
and we can immediately conclude that no anomalies exist at this order, 
since no functional $({\cal A}^{2}_{4})^{\a\b}$, with mass dimensions 
$+2$ and ghost charge $+4$ can be written, which satisfies the ghost 
equation \equ{6.7}.

At order $O(\theta^{2})$:
\eq
({\cal A}^{7}_{1})^{\a_{1}\b_{1}\a_{2}\b_{2}}
=
{\cal W}\
({\cal A}^{4}_{4})^{\a_{1}\b_{1}\a_{2}\b_{2}}\ .
\eqn{6.15}
There are three possibilities:
\begin{eqnarray}
({\cal A}^{4}_{4})^{\a_{1}\b_{1}\a_{2}\b_{2}}_{(1)}
&=&
\d^{\a_{1}\a_{2}}\d^{\b_{1}\b_{2}}\
T_{1}^{abcd}\
\int d^{3}x\
(\partial^{\m}c^{a})(\partial_{\m}c^{b})(\partial^{\n}c^{c})(\partial_{\n}c^{d})
\label{6.16}\\
({\cal A}^{4}_{4})^{\a_{1}\b_{1}\a_{2}\b_{2}}_{(2)}
&=&
\d^{\a_{1}\a_{2}}\
T_{2}^{abcd}\
\int d^{3}x\
(\partial^{\b_{1}}c^{a})(\partial^{\b_{2}}c^{b})(\partial^{\m}c^{c})(\partial_{\m}c^{d})
\label{6.17}\\
({\cal A}^{4}_{4})^{\a_{1}\b_{1}\a_{2}\b_{2}}_{(3)}
&=&
T_{3}^{abcd}\
\int d^{3}x\
(\partial^{\a_{1}}c^{a})
(\partial^{\a_{2}}c^{b})(\partial^{\b_{1}}c^{c})(\partial^{\b_{2}}c^{d})
\label{6.18}
\end{eqnarray}
where $T^{abcd}_{i}$ are 
invariant tensors. 

Due to the algebraic relation
\eq
\{B_{\Sigma},{\cal W}\}=0\ ,
\eqn{6.19}
which holds in the space of integrated functionals, the anomaly 
candidates are $B_{\Sigma}$-invariant. Nevertheless, it is immediate 
to recognize that, using again \equ{6.19}, none of the candidates belong 
to  the BRS cohomology, therefore, no $\theta$-dependent anomaly 
exists, at least at order $O(\theta^{2})$.

We could easily  carry on the analysis at higher orders in $\theta$, 
to finally prove that no $\theta$-dependent anomaly exists at all, 
thus leading to the nontrivial result that the noncommutative 
extension does not introduce anomalies, at least in Chern-Simons 
theory. But the aim of this section is just to give an example of the 
power of the result \equ{4.30}, namely the most general solution of the 
supersymmetry equation, which allows a much easier analysis of issues 
otherwise technically quite involved.

\section{Conclusions}

We have shown, on the example of the Chern-Simons model in three 
spacetime dimensions, that tackling the problem of its 
renormalizability by first solving for the linear vector supersymmetry, 
considerably shortens the algebraic work to be done; in 
particular, the absence of counterterms (finiteness) and of anomalies 
reduces just to a couple of lines.

The reason for this, is that the most general solution of the linear vector 
supersymmetry has the form \equ{4.30}, expressed by means of the 
operator ${\cal W}$ \equ{4.31}, which shifts the problem to a 
much smaller functional space , being characterized by a higher ghost 
charge, and zero dimensionality. 

We remark that, this property is related to the functional form of 
the operator $W_{\m}$, which is the same for any topological field 
theory, since it depends only on the gauge fixing part of the action 
and not on the particular model considered. 

It would be nice if this result could be extended to twisted 
topological field theories; here we have to face the problem that the 
vector supersymmetry is not linearly realized, and therefore the 
study of the cohomology of the $W_{\m}$ operators is not 
straightforward. In particular, for the twisted $N=2$ SYM model, it is known 
that the Lagrangian can be written as
\eq
\Sigma_{N=2SYM}={\cal W}X\ ,
\eqn{6.1}
but only modulo an exact BRS cocycle, and there are $W_{\m}$ 
invariant fields which suggest that the cohomology of the $W_{\m}$ 
operators might not be empty. Further investigations towards the 
generalization of our result to theories with nonlinear twisted 
supersymmetry are in progress.

What we have shown in this 
paper, is that any theory displaying a linear vector supersymmetry can be 
put in the form~\equ{6.1}, and hence shows relevant 
nonrenormalization properties.

Another point that we would like to stress, is that in the discussion 
of the renormalization of these supersymmetric models, it has not 
been necessary to solve any cohomological BRS condition, contrarily 
to what happens in ordinary cases. This is the 
simplification we mentioned more than once in this paper. 
Curiously enough, we here have a class of gauge field 
theories whose quantum extension is not determined by any BRS 
constraint. Nevertheless the gauge field theory nature of these 
models is apparent, being deeply encoded in the 
set of non gauge symmetries which are 
involved. We remind, indeed, that both the ghost equation \equ{2.14} and 
the supersymmetry  \equ{2.18} hold only if the Landau {\it gauge} is 
adopted. Moreover, for the classical action both the vector 
supersymmetry and the ghost equation are broken, so its determination 
is most easily and naturally done in terms of the standard BRS 
approach.

The hope is that a similar treatment, applied to noncommutative 
topological field theory models~\cite{Douglas:2001ba,Szabo:2001kg}, 
will help in the algebraic analysis 
of stability and anomaly. Indeed, the presence of the $\theta_{\m\n}$ 
parameter with negative power counting dimensions, raises very 
rapidly the dimensions of the field dependent breakings. Therefore, 
the possibility of lowering it by means of the solution of the vector 
supersymmetry looks like a promising way of attacking the problem.
Work is in progress in this direction.

%
%
%
%
%
%


\begin{thebibliography}{999}
\bibitem{Delduc:1989ft}
  F.~Delduc, F.~Gieres and S.~P.~Sorella,
  {\it Supersymmetry Of The D = 3 Chern-Simons Action In The Landau Gauge},
  Phys.\ Lett.\ B {\bf 225}, 367 (1989).
\bibitem{Maggiore:1991aa}
  N.~Maggiore and S.~P.~Sorella,
  {\it Finiteness of the topological models in the Landau gauge},
  Nucl.\ Phys.\ B {\bf 377}, 236 (1992).\bibitem{Alvarez-Gaume:1989wk}
  L.~Alvarez-Gaume, J.~M.~F.~Labastida and A.~V.~Ramallo,
  {\it A Note On Perturbative Chern-Simons Theory},
  Nucl.\ Phys.\ B {\bf 334}, 103 (1990).
\bibitem{Birmingham:1991ty}
  D.~Birmingham, M.~Blau, M.~Rakowski and G.~Thompson,
  {\it Topological field theory},
  Phys.\ Rept.\  {\bf 209}, 129 (1991).
\bibitem{Blasi:1990xz}
  A.~Blasi, O.~Piguet and S.~P.~Sorella,
  {\it Landau Gauge And Finiteness},
  Nucl.\ Phys.\ B {\bf 356}, 154 (1991).
\bibitem{Piguet:1995er}
  O.~Piguet and S.~P.~Sorella,
  {\it Algebraic renormalization: Perturbative renormalization, symmetries and
  anomalies},
  Lect.\ Notes Phys.\  {\bf M28}, 1 (1995).
\bibitem{West:1990tg}
  P.~C.~West,
  {\it Introduction To Supersymmetry And Supergravity},
   World Scientific (1990), Singapore.
\bibitem{Guadagnini:1990aw}
  E.~Guadagnini, N.~Maggiore and S.~P.~Sorella,
  {\it Supersymmetry Of The Three-Dimensional Einstein-Hilbert Gravity In The
  Landau Gauge},
  Phys.\ Lett.\ B {\bf 247}, 543 (1990).
\bibitem{Guadagnini:1990br}
  E.~Guadagnini, N.~Maggiore and S.~P.~Sorella,
  {\it Supersymmetric Structure Of Four-Dimensional Antisymmetric 
  Tensor Fields},
  Phys.\ Lett.\ B {\bf 255}, 65 (1991).
\bibitem{Maggiore:1992ug}
  N.~Maggiore and S.~P.~Sorella,
  {\it Perturbation theory for antisymmetric tensor fields in four-dimensions},
  Int.\ J.\ Mod.\ Phys.\ A {\bf 8}, 929 (1993)
  [arXiv:hep-th/9204044].
\bibitem{Blasi:1992hq}
  A.~Blasi and N.~Maggiore,
  {\it Infrared and ultraviolet finiteness of topological BF theory in
  two-dimensions},
  Class.\ Quant.\ Grav.\  {\bf 10}, 37 (1993)
  [arXiv:hep-th/9207008].
\bibitem{Birmingham:1991rh}
  D.~Birmingham and M.~Rakowski,
  {\it Vector supersymmetry in topological field theory},
  Phys.\ Lett.\ B {\bf 269}, 103 (1991).
\bibitem{Witten:1988ze}
  E.~Witten,
  {\it Topological Quantum Field Theory},
  Commun.\ Math.\ Phys.\  {\bf 117}, 353 (1988).
\bibitem{Fucito:1997xm}
  F.~Fucito, A.~Tanzini, L.~C.~Q.~Vilar, O.~S.~Ventura, C.~A.~G.~Sasaki and S.~P.~Sorella,
  {\it Algebraic renormalization: Perturbative twisted considerations on
  topological Yang-Mills theory and on N = 2 supersymmetric gauge  theories},
  arXiv:hep-th/9707209.
\bibitem{Blasi:2000qw}
  A.~Blasi, V.~E.~R.~Lemes, N.~Maggiore, S.~P.~Sorella, A.~Tanzini, O.~S.~Ventura and L.~C.~Q.~Vilar,
  {\it Perturbative beta function of N = 2 super Yang-Mills theories},
  JHEP {\bf 0005}, 039 (2000)
  [arXiv:hep-th/0004048].
\bibitem{Blasi:1989mw}
  A.~Blasi and R.~Collina,
  {\it Finiteness Of The Chern-Simons Model In Perturbation Theory},
  Nucl.\ Phys.\ B {\bf 345}, 472 (1990).
\bibitem{Delduc:1990je}
  F.~Delduc, C.~Lucchesi, O.~Piguet and S.~P.~Sorella,
  {\it Exact Scale Invariance Of The Chern-Simons Theory In The Landau Gauge},
  Nucl.\ Phys.\ B {\bf 346}, 313 (1990).
\bibitem{Lowenstein:1971vf}
  J.~H.~Lowenstein,
  {\it Normal Product Quantization Of Currents In Lagrangian Field Theory},
  Phys.\ Rev.\ D {\bf 4}, 2281 (1971).
\bibitem{Lowenstein:1971jk}
  J.~H.~Lowenstein,
  {\it Differential Vertex Operations In Lagrangian Field Theory},
  Commun.\ Math.\ Phys.\  {\bf 24}, 1 (1971).
\bibitem{Lam:1972mb}
  Y.~M.~Lam,
  {\it Perturbation Lagrangian Theory For Scalar Fields: Ward-Takahasi Identity And
  Current Algebra},
  Phys.\ Rev.\ D {\bf 6}, 2145 (1972).
\bibitem{Lam:1973qa}
  Y.~M.~Lam,
  {\it Equivalence Theorem On Bogolyubov-Parasiuk-Hepp-Zimmermann Renormalized
  Lagrangian Field Theories},
  Phys.\ Rev.\ D {\bf 7}, 2943 (1973).
\bibitem{Clark:1976ym}
  T.~E.~Clark and J.~H.~Lowenstein,
  {\it Generalization Of Zimmermann's Normal - Product Identity},
  Nucl.\ Phys.\ B {\bf 113}, 109 (1976).
\bibitem{Wess:1971yu}
  J.~Wess and B.~Zumino,
  {\it Consequences Of Anomalous Ward Identities},
  Phys.\ Lett.\ B {\bf 37}, 95 (1971).
\bibitem{Dixon:1991wi}
  J.~A.~Dixon,
  {\it Calculation of BRS cohomology with spectral sequences},
  Commun.\ Math.\ Phys.\  {\bf 139}, 495 (1991).
\bibitem{Zumino:1983ew}
  B.~Zumino,
  {\it Chiral Anomalies And Differential Geometry: 
  Lectures Given At Les Houches,
  August 1983}.
\bibitem{MacFarlane:1968vc}
  A.~J.~MacFarlane, A.~Sudbery and P.~H.~Weisz,
  {\it On Gell-Mann's Gamma Matrices, D Tensors And F Tensors, Octets, And
  Parametrizations Of SU(3)},
  Commun.\ Math.\ Phys.\  {\bf 11}, 77 (1968).
\bibitem{Douglas:2001ba}
  M.~R.~Douglas and N.~A.~Nekrasov,
  {\it Noncommutative field theory},
  Rev.\ Mod.\ Phys.\  {\bf 73}, 977 (2001)
  [arXiv:hep-th/0106048].
\bibitem{Szabo:2001kg}
  R.~J.~Szabo,
  {\it Quantum field theory on noncommutative spaces},
  Phys.\ Rept.\  {\bf 378}, 207 (2003)
  [arXiv:hep-th/0109162].
\bibitem{Blasi:2005bk}
  A.~Blasi, N.~Maggiore and M.~Montobbio,
  {\it Instabilities of noncommutative two dimensional BF model},
  Mod.\ Phys.\ Lett.\ A {\bf 20}, 2119 (2005)
  [arXiv:hep-th/0504218].
\bibitem{Blasi:2005vf}
  A.~Blasi, N.~Maggiore and M.~Montobbio,
  {\it Noncommutative two dimensional BF model},
  Nucl.\ Phys.\ B {\bf 740}, 281 (2006)
  [arXiv:hep-th/0512006].
\bibitem{Picariello:2001mu}
  M.~Picariello, A.~Quadri and S.~P.~Sorella,
  {\it Chern-Simons in the Seiberg-Witten map for non-commutative Abelian  gauge
  theories in 4D},
  JHEP {\bf 0201}, 045 (2002)
  [arXiv:hep-th/0110101].
\end{thebibliography}
\end{document}